# Robust Functional Magnetoencephalographic Brain Measures with 1.0 Millimeter Spatial Separation

**Don Krieger[1], Paul Shepard[2], Shawn Eagle[1], David O. Okonkwo[1]**

[1]Department of Neurological Surgery, University of Pittsburgh
[2]Department of Physics and Astronomy, University of Pittsburgh

Corresponding Author:
    Don Krieger, PhD
    Department of Neurological Surgery
    Suite 400B Presbyterian University Hospital
    200 Lothrop Street
    Pittsburgh, PA 15232
    kriegerd@upmc.edu
    +1(412)648-9654 office
    +1(412)521-4431 cell

**New & Noteworthy**

A global noninvasive survey of the human brain's functional neuroelectric activity including state-dependent silenced voxels is now obtainable with 1 mm differential resolution. Though comparable to local field potentials (LFPs) obtained with indwelling electrodes, MEG-derived neuroelectric currents are intermittent 80-msec fragments whereas LFP's are continuous. However, the coverage of LFPs is limited to the locations of the implanted electrodes whereas MEG-derived currents provide coverage of almost the entire brain including the white matter.

**Abstract**

Neuroelectric currents were extracted from free-running magnetoencephalographic (MEG) rest and task recordings in 617 normative subjects (ages: 18-87). State-dependent neuroelectric differential activation (DA) with spatial separation comparable to that of local field potentials was detected in more than 83% of this cohort.

Rest-high (rest greater than task) or task-high DA was found in more than 10,000 1-mm$^3$ voxels per subject in 77% of individual subjects. On average, 6% of the DA voxels bordered a second voxel whose DA was opposite; (i.e. one was rest-high and the other was task-high.). 516 subjects showed more than 100 such opposite voxel pairs 1 mm apart; 226 subjects showed more than 1000. The number of bordering voxel pairs with the same DA was consistently higher for almost all subjects and averaged 20%, ruling out the possibility that "opposite" bordering voxels occur simply by chance. For 65 brain regions, more than 10% of the cohort showed significantly more "same" than "opposite" pairs. These findings taken together support the conclusion that neuroelectric DA is consistently distinguishable at single 1 mm$^3$ brain voxels with 1-mm spatial separation.

When restricted to voxels with near-zero rest or task counts, significantly more rest-high than task-high voxels were found in 35 regions for at least 10% of the subjects. This inequality was not found when all DA-voxels were included. This supports the conclusion that the DA found in many rest-high voxels with near-zero task counts is due in part to task-dependent inhibition.

# Introduction

Local field potentials (LFPs) are detected using implanted electrode arrays. They are fluctuations in neural cell ensembles on a millimeter (mm) scale [1]. We report comparable resolution in distinguishing differences in activity between 1 mm$^3$ voxels 1 mm apart using noninvasive magnetoencephalographic (MEG) recordings in a large normative human lifespan cohort. The referee consensus solver [2-4] was used to identify, validate, and localize neuroelectric currents from MEG recordings ($p < 10^{-12}$ for each) . The solver is mathematically constrained to distinguish neuroelectric activity from one point on a 1 mm cubic grid compared with that at each of the 6 nearest neighbors. The work presented here tests and verifies the implied consequence of this constraint, i.e. the solver can resolve differences in neuroelectric activity with 1 mm separation. Neurophysiologic observations with this resolving power are presented which (a) suggest a more detailed and subject-specific localization of brain function than was previously available using noninvasive techniques and (b) provide evidence that task-dependent inhibition is detectable and is widely present. Specifically, we addressed the following research questions.

**Is differential activation (DA) detectable at single points on a 1-mm grid?** Neuroelectric currents were counted at each point on a 1-mm grid overlaid on the brain during rest and task conditions. An average of 668,187 points were tested per subject. The total counts found under the two conditions must be high enough to meet the stringent p-value threshold used to limit the rate of false discovery,

**Do voxels 1 mm apart show DA opposite to each other?** Though necessary, findings of significant DA-voxels are not sufficient to establish 1-mm differential resolution since one might actually be sampling the same functional volume when testing counts at grid points 1 mm apart. This possibility may be ruled out if pairs of grid points 1 mm apart are found which show DA in opposite directions. This approach was used in lieu of those which validate noninvasive functional imaging methods utilizing ground truth invasive measures, e.g. in vivo microelectrode arrays or optogenetics [5]. Coregistration with such methods would be problematic since the errors in localization accuracy obtainable with MEG do not, in general, provide 1 mm accuracy.

**Is local inhibition detectable during the task?** We hypothesized that local inhibition is more prominent during the task than during rest. If so, voxels with near-zero task counts would most likely be showing inhibition and we would find significantly more of them within a region than voxels with near-zero rest counts.

# Material and Methods

**Data analysis and glossary.**
**Spatial resolution** refers to the minimum separation at which functional differences may be reliably detected. This is synonymous with **differential spatial resolution**, **spatial separation**, and **resolving power**. **Localization accuracy/precision** is a second kind of **spatial resolution** which is not addressed by this work.

All data are presented as counts or the corresponding fractions. When assessing the significance of the difference between two counts within a brain volume, i.e. differential activation (**DA**), a **coin flip *p-value*** is computed. The great majority of these calculations are not used to test a null hypothesis but rather are used to screen the data to decide if the difference is unlikely enough to count or otherwise include the volume in further analysis.

A **voxel** is a volume of brain tissue defined by a 1-mm cubic grid which cover it. A **0-voxel** is the 1 mm$^3$ volume which contains a single grid point at its center. A **1-voxel** is the 7 mm$^2$ volume with six grid points 1 mm from a central point along the three coordinate axes ± 1 mm from the center. In general, all of the grid points in an **X-voxel** are within X mm of the central point.

A **DA-voxel** is a voxel for which the numbers of neuroelectric currents found during rest and task are significantly different from each other. A DA-volume is **task-high** or **rest-high** if the majority count is for task or rest respectively. Rest-high and task-high are equivalent to **task-low** and **rest-low** respectively. A **DA-region$_{subject}$** is a region for one subject for which the numbers of rest-high and task-high DA-voxels are significantly different from each other. A **DA-region$_{cohort}$** is a region for which the numbers of rest-high and task-high DA-region$_{subjects}$ are significantly different from each other.

The thresholds for significance to decide DA were selected to ensure that the rates of false discovery were held near zero. For the primary tests of 1-mm resolution, several hundred thousand DA-voxels were tested for each subject. A sliding threshold was used (see Figure 1) ranging from p < 10$^{-6}$ for voxels with as few as 21 currents to 10$^{-15}$ for voxels with 1000 currents to 10$^{-27}$ for voxels with 5000 currents. For tests by region, 50,000 or fewer voxels were tested per region and a fixed threshold of 10$^{-7}$ was used. For both DA-region$_{subjects}$ and DA-region$_{cohorts}$, 2784 tests were run (174 regions x 16 voxel sizes) and a fixed threshold of 10$^{-4}$ was used.

p(n,m,r) is the **coin-flip *p-value*** for getting at least m$_\uparrow$ heads in n coin flips where m$_\uparrow$ is the greater of the two counts, m and (n-m). r is the weight of the coin, 0.5 if the coin is unbiased. p(n,m,r) is intuitive, accurate, and readily calculated for any r, $0.0 \leq r \leq 1.0$.

$$p(n, m, r) = \sum_{k=m_\uparrow,n} C(n,k) r^k (1-r)^{n-k} \qquad (1)$$

Each term in the sum is the chance that exactly k of n tosses will be heads. This can happen in any of C(n,k) ways where $C(n,k) = \frac{n!}{k!\,(n-k)!}$. The chance for each of the C(n,k) possibilities is $r^k(1-r)^{n-k}$.

**Data source.**

Resting and task magnetoencephalographic (MEG) recordings from the same sitting (*n* = 617), a high resolution T1 weighted MRI scan (*n* = 617, ages 18-87), and diffusion weighted MRI scan (*n* =587) were provided by the Cambridge Centre for Ageing and Neuroscience (Cam-CAN) Normative Data Repository [6,7]. The MEG recordings include 560 seconds of eyes-closed rest and 560 seconds of eyes-open sensorimotor task performance [6]. For the sensorimotor task, subjects detected visual and auditory stimuli and responded to detection of each with a button press with the right index finger. For all recordings, head position within the MEG scanner array was monitored continuously and the forward solution used by the solver was corrected for head movement once per second.

The raw MEG data from each subject were transformed to a collection of probabilistically validated neuroelectric currents using the referee consensus solver [2-4]. Each current was identified, localized, and then validated (p < 10$^{-12}$) using a gradient search which is conceptually similar to that used in equivalent current dipole source localization [8]. The solver is capable of reliably and accurately identifying one neuroelectric current at a time, regardless of the number of currents which are present. This property enables efficient deployment on a wide variety of supercomputing resources as a scavenger which utilizes backfill compute cycles which would otherwise be wasted [2]. The primary data processing step required more than 40,000,000 processor hours on the Open Science Grid [9,10].

The average current count per subject per minute exceeded 500,000. Neuroelectric current counts for each subject were tallied separately from the 560-second rest and task recordings for each of 68 cortical regions, 68 adjacent rims of white matter up to 5 mm thick, 18 deep white matter tracts, and 19 subcortical regions. The xyz coordinates of the volumes which contain the deep white matter tracts were identified with Tracula, version 1.22 [11]. Those which contain the subcortical and cortical regions and their adjacent white matter rims were identified with Freesurfer version 5.3 using its default Desikan-Killiany atlas parcellation [12,13].

**Voxel identification and neuroelectric current counting.**
Voxels of several sizes confined to each region were identified and counts tallied for them as follows. The solver identifies and validates neuroelectric currents on a 1 mm grid which covers the brain. Hence the smallest voxel size used in the analysis is a single point on the grid. This is called a 0-voxel. An N-voxel includes grid points within N mm of the central point. The average number of 0-voxels per subject was 668,187, i.e. currents were found in about half the average brain volume (1,376,850 mm$^3$).

For each region within a single subject, a 0-quintuple was compiled for each 0-voxel, (x,y,z,$^0$count$_{rest}$,$^0$count$_{task}$), where x, y, and z are the coordinates of the central grid point and $^0$count$_{rest}$ and $^0$count$_{task}$ were the neuroelectric current counts found at that point during rest and task. To generate a list of N-quintuples for larger voxels, e.g. 1-voxel, $^1$count$_{rest}$ and $^1$count$_{task}$ and the corresponding 1-quintuples were determined as follows. The first 0-quintuple was placed in the 1-quintuple list. For each of the remaining 0-quintuples within the region, e.g. the n$^{th}$, the Euclidean distance to each 1-quintuple was computed and the counts were added to the first one found for which the distance ≤ 1.0 mm. If no such 1-quintuple was found, a new 1-quintuple was added to the list with its counts and with the xyz coordinates from the n$^{th}$ 0-quintuple as its central point. Suppose the n$^{th}$ 0-quintuple was 2.0 mm from the nearest 1-mm voxel, **K**. Then it became the center point of the next 1-quintuple, **L**. Note that if a 0-quintuple later in the list fell between the two center points, it was added to **K** because **K** came earlier in the list.

Quintuple lists were generated for Euclidean distances of 0, 1, …, 12 mm and for $\sqrt{2}$, $\sqrt{3}$, and 2.5 mm. As stated above, 0-voxels contain a single point. 1-voxels include a maximum of 6 xyz locations adjacent to the central location along one of the xyz axes. The $\sqrt{2}$-voxels include up to 12 additional locations which are diagonally adjacent to the central location and in the xy, yz, or xz plane with it. $\sqrt{3}$-voxels include up to 8 additional locations which are diagonally adjacent to the center at the far corners. Only the 0- and $\sqrt{3}$-voxels may contain all and only the grid points in a cubic voxel. As the voxel size increases, the voxel shape more nearly approximates a sphere, but only grid points within a single region may be included in a voxel. Example regional voxel counts over all 617 subjects are as follows. For the left lateral occipital cortex, the medians were 6,960 (0-voxel), 3,024 (1-voxel), 904 (2-voxel), and 223 (4-voxel). For the left superior frontal cortex, the medians were 8,630, 3,718, 1,153, and 293, and for the brain stem they were 11,373, 4,700, 1,165, and 313.

**Bordering voxels.**
The distance below which two voxels may border each other is defined as the integer 1 mm larger than the size of the voxels. For example, for 0-voxels and 1-voxels, the distances are 1 and 2 mm respectively; for 2-voxels and 2.5-voxels, the distance is 3 mm. Consider two fully populated 1-voxels with central points (0,0,0) and (3,0,0). They are adjacent but do not "border" since they are 3 mm apart. But 1-voxels with centers at (1,1,0) and (1,1,1) would border (0,0,0).

If a 1-voxel existed with central point (0,2,0), it would border (0,0,0) but could not be fully populated.

**(1) Is DA detectable at single points on a 1-mm grid?**

The counts found within a region for rest and task are $^{region}count_{rest}$ and $^{region}count_{task}$ respectively. The regional resting and task count fractions are $\Phi_{rest}$ and $\Phi_{task}$ where $\Phi_{rest} + \Phi_{task} = 1.0$ and $\Phi_{rest} = {}^{region}count_{rest}/({}^{region}count_{rest} + {}^{region}count_{task})$. These are the expected fractions for each of the voxels in the region that have no DA. Over all subjects x regions, the *mean* and *standard deviation* for these expected fractions were 0.509 and 0.071 respectively with the great majority ranging from 0.42 – 0.59. Expected fractions obtained in this way control well for differences in counts due to differences in (a) recording times and (b) presence of artifacts in the recordings. However, since expected fractions typically differed from one region to the next, bordering DA-voxels were counted only if both voxels were in the same region.

To assess DA for a voxel in a particular region, a coin-flip *p-value* is computed by comparing $^{region}count_{rest}$ and $^{region}count_{task}$ to $^{voxel}count_{rest}$ and $^{voxel}count_{task}$. The voxel count fractions are $\delta_{rest} + \delta_{task} = 1.0$ where $\delta_{rest} = {}^{voxel}count_{rest}/({}^{voxel}count_{rest} + {}^{voxel}count_{task})$. Without loss of generalization, assume that $\delta_{rest} \geq \Phi_{rest}$ and therefore $\delta_{task} \leq \Phi_{task}$. Then $p(n,m,r)$ is computed as in equation (1) above where $^{voxel}count_{rest} = m_\uparrow$, $\Phi_{rest} = r$, $^{voxel}count_{rest} + {}^{voxel}count_{task} = n$.

The lower panel of figure 1 is a scatter plot showing n (x-axis) and *p-value* (y-axis) for each voxel for subject 331. Note that all size voxels are included. The set of voxels bounding the left side of the plot about the line with slope -0.3 is labelled "Optimal *p*-value." These voxel values arise from the dependence of the best achievable *p*-value, $p_{opt}$ on n. Suppose n = 10. $P_{opt}$ is obtained when one gets 10 heads out of 10 flips. If the coin is "fair," i.e. the chance of getting heads is 0.5, $p = 0.5^{10} = 1/1024 \approx 10^{-3}$. If n is 20, $p_{opt} = 0.5^{20} = 1/1,048,576 \approx 10^{-6}$. Log($p_{opt}$) decreases by 3.0 when the number of flips increases by 10, i.e. the equation $\log(p_{opt}) = -0.3n$ has slope -0.3.

Insert Figure 1

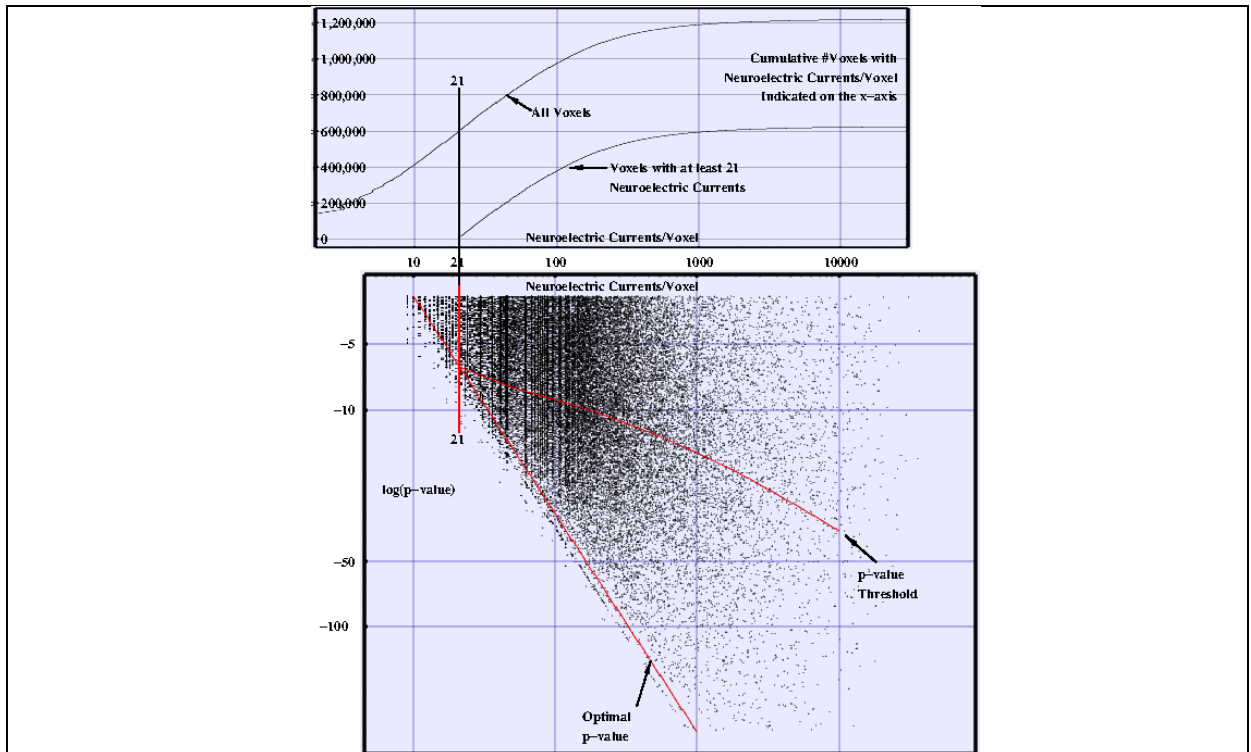

**Figure 1: Differentially activated (DA) voxels.** Neuroelectric current counts for resting and task recordings were compared within each voxel for each subject. Results for one typical subject are shown including voxels with all sizes. The lower panel shows a typical scatter plot for one subject. For each voxel, the x-axis is the sum of the neuroelectric currents from both resting and task recordings; the y-axis is log(p-value) for the difference between resting and task recordings. Voxels for which log(p-value) > -3.0 are excluded. The line labelled "Optimal p-value" indicates the locations in the plot of voxels for which the differential is maximized, e.g. observed resting and task counts = 40, 0; expected counts = 20, 20. The observations show some spread about the line because the expected ratio varies about 50/50. The line labelled "p-value threshold" is the line above which voxel p-values do not show sufficient significance to be included for further analysis. The upper panel shows two cumulative histograms for the x-axis variable, i.e. total neuroelectric current counts within each voxel. The upper trace shows the histogram for all voxels; the lower trace shows all voxels with total counts ≥ 21. 21 is the minimum count for which a 2-sided *p*-value < $10^{-6}$ can be achieved. Only these were tested for DA. See the Methods text for more details.

The upper panel of the figure is a cumulative histogram showing the number of voxels covering this subject's brain (y-axis) with at least the neuroelectric current count listed on the x-axis. DA within each region x voxel size was tested for those with at least 21 rest+task counts, i.e. about 600,000 (lower trace). To control for "false discovery" with so many tests, the threshold to accept a voxel as showing DA was set as follows. Since $p_{opt}$ and statistical power depend monotonically on n, the threshold *p-value* for each voxel is set to be progressively more stringent for voxels with as n increases:

$$^{threshold}p\text{-}value_n = -6.0 - \left(0.3 \times \sqrt{n_{voxel} - 21}\right) \quad (2)$$

The curve described by this equation is shown in the figure and labelled "*p*-value Threshold." For voxels with the minimum testable n per equation (2), 21, the threshold p-value is $10^{-6}$ which is only achievable if the 21 counts within the voxels are all for one condition, i.e. either 21 rest and 0 task or 0 rest and 21 task. The upper panel of the figures shows that this threshold count reduces the number of voxels to be tested to about 600,000. The progressively more stringent *p*-value threshold with increasing n ensures that false discovery rate is held near zero for all

600,000+ tests, i.e. almost all of the voxels which show significance are likely differentially activated. That number is in excess of 13,000 for this typical subject.

Insert Figure 2

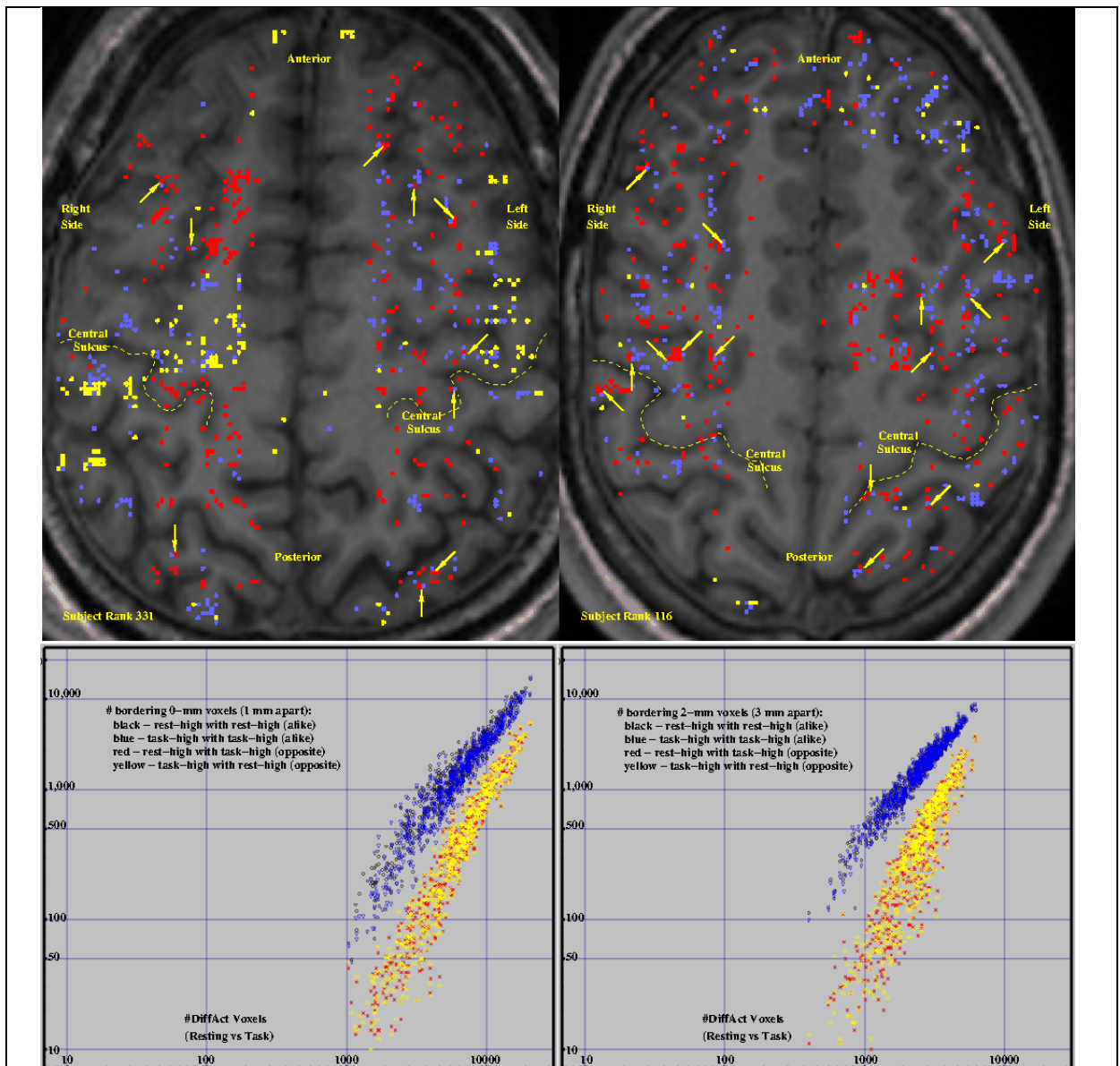

**Figure 2. Confirmation of 1 mm resolution – two typical subjects (upper panels) and cohort-wide (lower panels).** (Upper Panels) Neuroelectric current counts for resting and task recordings were compared within each 0-mm voxel for each subject separately. Typical DA-voxels (p < $10^{-6}$ for each) from two subject are shown superimposed on these 1-mm MRI slices about 25 mm below the apex of the brain. The central sulci provide landmarks and are labelled and indicated with dotted lines. Voxels are shown for which resting vs task DA was significant using the criteria described above and in Figure 1. The subjects with the 331st and 116th highest number of DA- voxels are shown. Task-high voxels ($count_{task} \gg count_{rest}$) are shown in red. Rest-high voxels are shown in blue or yellow for task count = 0. Yellow arrows highlight voxel pairs with opposite DA which border each other, i.e. are 1 mm apart.

(Lower Panels) The number of bordering voxels for each subject (y-axis) is plotted against the total number of DA-voxels for the subject (x-axis). Voxels which border two or more others were counted only once. Alike voxel pair counts (rest-high|rest-high and task-high|task-high) are plotted in black and blue. Opposite counts (rest-high|task-high) are plotted in red and yellow. Results for 0-mm voxels (left panel) and 2-mm voxels (right panel) are shown. Subjects were excluded for whom the number of opposite voxel pairs was less than 10. For 0-mm voxels, 598 were included; for 2-mm voxels, 587 were included respectively. See the Methods and Results text for more details.

**(2) Do voxels 1 mm apart show DA opposite to each other?**

In Figure 2, the upper panels show a sampling of DA 0-voxels from two typical subjects in an MRI slice about 25 mm below the apex of the brain. The total number of opposite bordering DA-voxel pairs found were 613 (subject 331) and 1616 (subject 116). This provides a "yes" answer to question 2 for these two subjects. Note that all bordering pairs were constrained to be within the same region to avoid artifactually identified "opposite" pairs due to differences in expected count fractions from one region to the next.

For a more conclusive answer, the 0-voxels which border at least one other 0-voxel with opposite DA were counted for every subject and tallies were plotted in yellow and red in Figure 2, lower left panel (*mean*: 1067). Note that the counts for alike 0-voxel bordering pairs were consistently higher (blue and black dots in the Figure). The following procedures was used to ensure that the counts may be directly compared. For each subject, the $\rho$ rest-high voxels and $\tau$ task-high voxels were located at coordinate triplets $^R(x,y,z)_{i=1,\rho}$ and $^T(x,y,z)_{j=1,\tau}$ respectively. Note that each of the $^R(x,y,z)$'s and $^T(x,y,z)$'s are unique. We presume without loss of generality that $\rho \leq \tau$. We draw randomly from the $^R(x,y,z)$'s to obtain $^{R1}(x,y,z)$ and $^{R2}(x,y,z)$, each with $\rho/2$ voxels. We also draw randomly from the $^T(x,y,z)$'s to obtain $^{T1}(x,y,z)$ and $^{T2}(x,y,z)$, each also with $\rho/2$ voxels. $\rho/2$ is the x-axis variable in Figure 2 (lower panels) and is labelled: #DiffAct Voxels. This reduction of the number of voxels to half the lesser of $\rho$ and $\tau$ provides an unbiased comparison of alike DA-voxel pairs with opposite pairs.

## Results and Discussion

**Is DA detectable at 0-mm voxels, i.e. single points on a 1-mm grid?**

The mean number of DA 0-mm voxels per subject over the Cam-CAN cohort was 14,128. That is 2.1% of the mean of 668,187 voxels per subject at which neuroelectric counts were found. Of the 617 subjects in the cohort, 391 showed 10,000 or more, 536 showed 5000 or more, and 8 showed 1000 or less. The cohort-wide answer to this critical capability in achieving 1-mm resolution is "yes."

**Do grid points occur which show DA opposite to those 1 mm away?**

The mean number of DA 0-mm voxels per subject found within 1 mm of another voxel with opposite DA was 1067. Subjects 331 and 116 (Figure 2, Upper Panel) showed 613 and 1616 respectively. Of the 617 subjects in the cohort (Figure 2, Lower Left Panel), 226 showed 1000 or more, 516 showed 100 or more, and 20 showed 10 or less. The cohort-wide answer to this second critical capability in achieving 1-mm resolution is also "yes."

Figure 2 shows that the counts for alike voxel pairs is consistently greater than that for opposite pairs. These counts were tallied for each subject x region x voxel size and compared with coin flip p-values. If rest-high and task-high DA-voxels were randomly distributed in space, these counts would be nearly equal. Figure 3 shows that they are consistently unequal in the expected direction, i.e. same-high, for many regions cohort-wide. This result is consistently strongest for 0-voxels, i.e. 1-mm$^3$ voxel pairs which are 1 mm apart.

Insert Figure 3

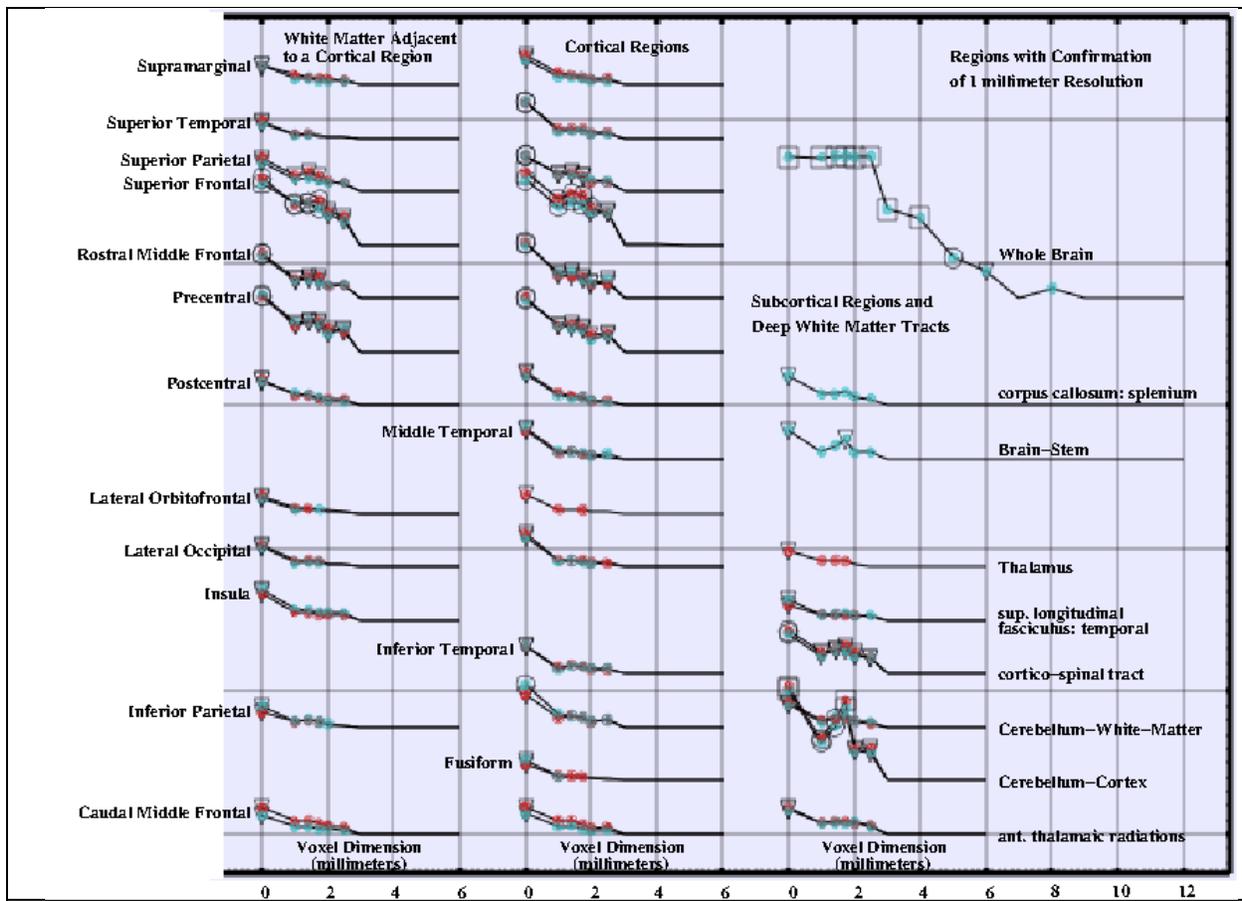

**Figure 3. Confirmation of 1 mm resolution by region, cohort-wide.** For each region x voxel size, the subjects with a significant difference in alike vs opposite bordering voxels were tallied. Only alike-high significance was found, as expected. Traces are shown for regions with significance for at least 0-voxels ($p < 10^{-4}$ for each). The y-axis is the fraction of the cohort for which the region is alike-high. The total number of same-high subjects is indicated with symbols. No same-high subjects were found. triangle: > 60 (10% of the cohort); circle: > 150 (25%); square: > 300. Red dots indicate significance for left side regions, cyan for midline and right side regions. See the Methods and Results text for more details.

**Is local inhibition detectable during the task?**

When significant DA is found within a voxel, e.g. rest-high, it isn't possible to determine to what extent excitation during rest and inhibition during task contribute. However, very low task counts suggest the presence of task-dependent inhibition. We conjecture that task-dependent inhibition, i.e. task-low DA-voxels, are more common than rest-dependent inhibition, i.e. rest-low DA-voxels. All DA-voxels with 0-9 minority counts were tallied for each subject x region x voxel size and compared using coin flip p-values. Figure 4 shows that the counts for task-low DA-voxels are significantly high for many regions cohort-wide. Tallies were also obtained for 0-voxels with 0 rest or task counts. This produced consistently stronger findings as shown in the Figure. Finally, tallies were obtained using all DA-voxels; this produced much weaker findings, also as expected since larger minority counts are less suggestive of inhibition. These results suggest that local inhibition is detectable during the task and is widely present.

Insert Figure 4

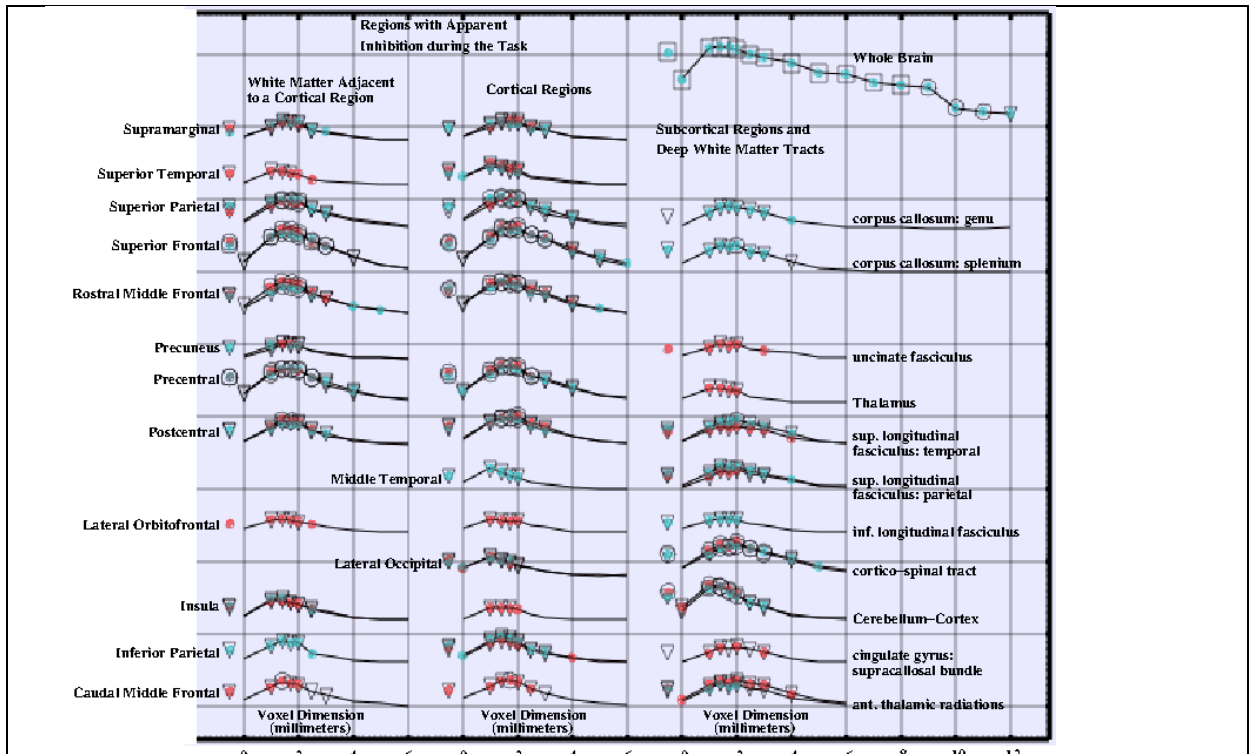

**Figure 4. Inhibition during the task by region, cohort-wide.** For each region, the numbers of subjects with majority rest-high voxels or majority task-high voxels were compared. DA-voxels were counted only with 0-9 currents for either rest or task. Regions are included for which DA-region$_{cohort}$ was found for at least 3 voxel sizes (p < $10^{-4}$ for each). The x-axis is voxel dimension. The y-axis is the fraction of the cohort for which the region is rest-high. The total number of rest-low + task-low subjects is indicated with symbols: triangle: > 60 (10% of the cohort); circle: > 150 (25%); square: > 300. Red dots indicate significance for left side regions, cyan for midline and right side regions.

All significant DA-region$_{cohort}$ were task-low, i.e. only rest-high DA-region$_{cohort}$'s were found, as expected. Results for 0-voxels with zero task currents are shown to the left of each trace; they are consistently stronger than for 0-voxels with 0-9 task counts, as expected. Results for tallies using all DA-voxels were weaker still. Of the 2768 tests run for every region by voxel size, 369 DA-region$_{cohort}$'s were significant for voxels with 0-9 task counts whereas only 29 were significant when all DA-voxels were included. See the Methods and Results text for more details.

**Functional Unit Size.**

For both bordering DA-voxels by region (Figure 3) and task-dependent inhibition by region (Figure 4), cohort-wide results are shown for a wide range of voxel volumes. Distinct maxima are seen in the tracings for most regions for smaller voxels. For bordering voxels, we conjecture that a local peak for a region at a particular voxel volume suggests that that voxel volume is a preferred functional unit size within the region. We define a functional unit empirically as a brain volume within which the neural elements predominantly cooperate or otherwise demonstrate a common behavior during rest or task. We reason that the all pairs must straddle the boundary between a rest-high and task-high functional unit. The likelihood of finding such pairs is maximized when the region is populated by units with that same volume.

**Limitations.**

The MEG recordings are processed in 80-msec segments. For each segment, the solver is limited to identifying at most one neuroelectric current for each 8x8x8 mm voxel. This limitation is imposed in recognition of the limited information provided in an 80-msec recording from 306 magnetic field sensors sampled at 1000 points/sec. Given this limit, the number of currents identified for each data fragment typically ranges no higher than 1500. Though large in comparison with the number of currents that are obtainable with other current extraction methods, 1500 is small compared to the 1,000,000+ mm$^3$ volume of the brain covered by the solver.

Identification of simultaneously occurring currents within 1 mm of each other is rare as is identification of currents at the same location from consecutive 80-msec segments. Hence the extracted 80-msec waveforms for each neuroelectric current are fragments rather than continuous. Though comparable in resolution to local field potentials (LFPs) obtained with direct recording electrodes, MEG-derived neuroelectric currents are intermittent 80-msec fragments whereas LFP's have the advantage that they are continuous. On the other hand, the coverage of LFPs is limited to the locations of the implanted electrodes whereas MEG-derived currents provide coverage of almost the entire brain including the white matter. [4]

In comparison with the spatially differential resolution shown here, localization accuracy is likely not as good since it may be degraded by (a) errors in the spatial precision of the head model and forward solution used in the solver's search for neuroelectric currents and (b) errors in the static coregistration of the subject's head position in the MEG and MRI scanners. As mentioned above, for all recordings, head position within the MEG scanner array was monitored continuously and the forward solution used by the solver was corrected for head position once per second.

**Concluding statement.**

A global noninvasive survey of the human brain's functional neuroelectric activity including state-dependent silenced voxels is now obtainable with 1 mm differential resolution.

## Acknowledgments:

We gratefully acknowledge the invaluable contributions made to this effort by the Cambridge (UK) Centre for Ageing and Neuroscience (Cam-CAN), the Extreme Science and Engineering Development Environment (XSede, IBN130001), The Texas Advanced Computing Center (TACC, TG-IBN130001), the Open Science Grid (OSG), the San Diego Supercomputing Center, the Pittsburgh Supercomputing Center, the XSede Neuroscience Gateway, Grace Jane Gollinger, Darren Price, Ethan Knights, Mats Rynge, Rob Gardner, Frank Wurthwein, Derek Simmel, Mahidhar Tatineni, Data used in the preparation of this work were obtained from the CamCAN repository [6,7]. The OSG [9,10] is supported by the National Science Foundation, 1148698, and the US Department of Energy's Office of Science.